\documentclass[12pt]{article}

\usepackage{mathrsfs}

\usepackage{amsmath,amssymb}
\usepackage{amsmath}
   \usepackage{amscd}
   \usepackage{amsfonts}
   \usepackage{amssymb}
   \usepackage{amsthm}
   \usepackage{epsfig}

\def\be{\begin{equation}}
\def\ee{\end{equation}}
\def\ba{\begin{eqnarray}}
\def\ea{\end{eqnarray}}

\def\ket[#1]{\left|#1\right>}
\def\bra[#1]{\left<#1\right|}

\def\Zop{\bbbz}
\def\bbbz {{\sf Z\!\!Z}}

\def\dg{\dagger}
\def\a{\alpha}

\def\b{\beta}
\def\l{\lambda}

\def\p{\partial}

\def\Str{{\rm  Str}}
\def\ket[#1]{\left|#1\right>}
\def\bra[#1]{\left<#1\right|}
\def\ie{{\it i.e.}}

\parskip=\medskipamount

\input epsf
\usepackage{epsfig,color,latexsym}

\renewcommand{\theequation}{\thesection.\arabic{equation}}

\begin{document}
\renewcommand{\theequation}{\thesection.\arabic{equation}}

\thispagestyle{empty}
\def\thefootnote{\fnsymbol{footnote}}\begin{flushright}
MIT-CTP-nnnn\\
Imperial/TP/2-08/nn
\end{flushright}\vskip 0.5cm\begin{center}
\LARGE{\bf  Green-Schwarz action for Type IIA strings on $AdS_4\times CP^3$}
\end{center}\vskip 0.8cm\begin{center}{\large
B. Stefa\'nski, jr.$^{1,2}$
}
\vskip 0.2cm{\it $^1$ Center for Theoretical Physics \\ Laboratory for Nuclear Science,
\\ Massachusetts Institute of Technology \\  Cambridge, MA  02139, USA}
\vskip 0.2cm{\it $^2$ Theoretical Physics Group,  Blackett  Laboratory, \\
Imperial College,\\ London SW7 2BZ, U.K.}
\end{center}
\vskip 1.0cm
\begin{abstract}\noindent
We present the Green-Schwarz action for Type IIA strings on $AdS_4\times CP^3$. The action is based on a
$\Zop_4$ automorphism of the coset $OSp(4|6)/(SO(1,3)\times SU(3)\times U(1))$. The equations of motion admit a representation 
in terms of a Lax connection, showing that the system is classically integrable.
\end{abstract}

\vfill

\setcounter{footnote}{0}
\def\thefootnote{\arabic{footnote}}
\newpage

\renewcommand{\theequation}{\thesection.\arabic{equation}}

\section{Introduction}
\setcounter{equation}{0}

Following the papers of Bagger and Lambert~\cite{bl} (see also~\cite{gust}) there has been 
much interest in ${\cal N}=8$ $d=3$ supersymmetric theories.~\footnote{Earlier work on such theories with less supersymmetry can be found 
in~\cite{earlier} and references therein.} Recently it has been suggested that 
certain ${\cal N}=6$ $D=2+1$ $U(N)\times U(N)$ Chern-Simons theories at
level $k$ with matter have a dual description as M-theory on $AdS_4\times S^7/\Zop_k$~\cite{m}. We may view
the $S^7$ as a Hopf fibration with the radius of $CP^3$ proportional to $Nk$ and the radius of the Hopf circle
proportional to $N^{1/6}k^{-5/6}$. The authors of~\cite{m} then identify a limit of the $N$-$k$ parameter space
in which the M-theory is described by a weakly coupled string theory on a $AdS_4\times CP^3$ background
\be
ds^2_{\mbox{str}}=\frac{R^3}{k}(\frac{1}{4}ds^2_{AdS_4}+ds^2_{CP^3})\,,\label{bkd}
\ee
together with a four-form RR flux on $AdS_4$ and a two-form RR flux on $CP^1\subset CP^3$. In this note we
propose a GS action for type IIA string theory~(\ref{bkd}). We follow the approach developed in a series of papers 
in the study of GS actions in Minkowski and $AdS_5\times S^5$ spacetimes~\cite{MT}. 
This approach has proven to be particularly useful in the studies of integrability of
string theory; indeed we are able to construct a Lax connection for this action in analogy with the results
of~\cite{BPR}. In section 2 we present an explicit realisation of the 
$\Zop_4$ automorphism and in section 3 
we present the action, discuss $\kappa$-symmetry for the model and construct the Lax pair for it; 
conclusions are presented in section 4.

\section{A $\Zop_4$ automorphism}
\setcounter{equation}{0}

The GS action presented in this paper will be based on the super-coset
\be
\frac{OSp(4|6)}{SO(1,3)\times SU(3)\times U(1)}\,.
\label{coset}
\ee
The Lie super-algebra can be represented by a super-matrix $M$ written in block-matrix form
\be
M=\left(\begin{matrix} A &B \\ C & D\end{matrix}\right)\,,
\ee
where $A$ is $4\times 4$, $B$ is $4\times 6$ $C$ is $6\times 4$ and $D$ is 
$6\times 6$. In order to belong to the super-algebra $OSp(4|6)$ we require 
that $M$ satisfies
\be
\Omega M+ M^t\Omega=0\,,
\ee
where the supertranspose of $M$ is given by
\be
M^t=\left(\begin{matrix} A^t &C^t \\ -B^t & D^t\end{matrix}\right)\,,
\ee
while the fixed matrix $\Omega$ is
\be
\Omega =\left(\begin{matrix} \omega & 0 \\ 0 & 1 \end{matrix} \right)\,,
\ee
and
\be
\omega = \left(\begin{matrix} 0 & 1 & 0 & 0 \\ -1 & 0 & 0 & 0 \\ 0&0&0&1 
\\
0&0&-1&0\end{matrix}\right)\,.
\ee
In terms of the block matrices this reduces to 
\be
A^t\omega+\omega A =0\,,\qquad D+D^t=0\,,\qquad C=B^t\omega\,.
\label{cptsualg}
\ee
In particular  $A$ is a $Sp(4)$ matrix, $D$ is a $SO(6)$ matrix. 
We also require the reality condition
\be
M^\dagger=-\left(\begin{matrix} C^{-1}&0 \\ 0& 1\end{matrix}\right) 
M \left(\begin{matrix} C&0 \\ 0& 1\end{matrix}\right)\,,
\label{realcond}
\ee
with $C$ defined in appendix~\ref{appa}. The bosonic sub-algebra of $M$ then 
generates $Sp(2,R)\times SO(6)\sim SO(2,3)\times SU(4)$ (see appendix~\ref{appa} for more detail).

Next we define the following automorphism
\be
G(M)\equiv g^{-1}Mg\equiv
\left(\begin{matrix} K& 0 \\ 0 &
-iL\end{matrix}\right) M \left(\begin{matrix} K& 0 \\ 0 &
iL\end{matrix}\right)
=
\left(\begin{matrix} KAK & iK B L \\ -iL C K & 
LDL\end{matrix}\right)\,,
\ee
\label{automorphism}
where
\be
K=\left(\begin{matrix} 1_2 & 0 \\ 0 & -1_2\end{matrix}\right)\,,
\ee
and
\be
L = \left(\begin{matrix} 0 & i & 0 & 0 & 0 & 0 \\ -i & 0 & 0 & 0 & 0 & 0 
\\ 0&0&0&i&0&0\\
0&0&-i&0&0&0 \\ 0&0&0&0&0&i\\
0&0&0&0&-i&0
\end{matrix}\right)\,.
\ee
It is easy to see that  $G(M)\in OSp(4|6)$ since
\be
K\omega=\omega K\,,\qquad K^t=K\,,\qquad L^t=-L\,,\qquad K^2=1_4\,,\qquad 
L^2=1_6\,,
\ee
and so me may verify equation~(\ref{cptsualg}) explicitly
\ba
(KAK)^t\omega+\omega KAK =K(A^t\omega +\omega A)K &=& 0\,,\\
(LDL)^t+LDL = L(D^t+D)L &=& 0\,,\\
(iK B L)^t\omega = -iLB^t\omega K &=& -iLC K \,.
\ea
It is also easy to show that
\be
G^2(M)=\left(\begin{matrix} A & -B \\ -C & D\end{matrix}\right)\,.
\ee
which shows that $G$ generates a $\Zop_4$ action on $OSp(4|6)$. Since 
$G(M)=g^{-1}Mg$ this is an automorphism of the Lie superalgebra, in other 
words
\be
\left[G(M_1)\,,\,G(M_2)\right\}=G(\left[M_1\,,\,M_2\right\})\,.
\ee
Notice also that while $iL\in SO(6)$, $K\not\in Sp(4)$, and so $g\not\in 
OSp(4|6)$; as a result $G$ is an outer automorphism. As shown in appendix~\ref{appa}
 the sub-algebra left invariant by $G$ is
$SO(1,3)\times SU(3)\times U(1)$.

\section{GS action for Type IIA string theory on $AdS_4\times CP^3$}
\setcounter{equation}{0}

In this section we construct the spacetime supersymmetric action for Type IIA string theory on $AdS_4\times 
CP^3$. We first present the form of the action then discuss $\kappa$-symmetry and finally show that the action 
is supersymmetric.

\subsection{The action}

Let us briefly recall the construction of the GS action on a super-coset
$G/H$. We require that: (i) $H$ be bosonic and, (ii) $G$ admit a $\Zop_4$ 
automorphism
that leaves $H$ invariant, acts by $-1$ on the remaining bosonic part of 
$G/H$, and by $\pm i$ on the
fermionic part of $G/H$. The currents
$j_\alpha=g^\dg\p_\alpha g$ can then be decomposed as
\be
j_\alpha=j^{(0)}_\alpha+j^{(1)}_\alpha+j^{(2)}_\alpha+j^{(3)}_\alpha\,,\label{currz4dec}
\ee
where $j^{(k)}$ has eigenvalue $i^k$ under the $\Zop_4$ automorphism. The 
$+1$ eigenspace is
\be
M^{(0)}=\left\{\left(\begin{matrix} A^{(0)} &0 \\ 0 & D^{(0)}\end{matrix}\right) | A^{(0)}=KA^{(0)}K\,,D^{(0)}=LD^{(0)}L\right\}\,,
\ee
where $A^{(0)}=KA^{(0)}K$ implies
\be
A^{(0)}=\left(\begin{matrix} a_1 &a_2 \\ a_3&-a_1\end{matrix}\right)\oplus
\left(\begin{matrix} a_4 &a_5  \\ a_6&-a_4
\end{matrix}\right)\sim SO(1,3)\,,
\ee
where $A^{(0)}$ satisfies the reality condition~(\ref{realcond}) and hence spans $SO(1,3)$ as shown in 
appendix A.
Similarly we may describe the solutions of $D=LDL$ in terms of linear combination of anti-symmetric matrices 
$(M^{ij})_{kl}=\delta_{ik}\delta_{jl}-\delta_{il}\delta_{jk}$. We find the condition $D=LDL$ implies
\ba
D^{(0)}&=&\left\{M^{12}\,,M^{34}\,,M^{56}\,,M_{15}\mp iM_{16} \pm iM_{25}+M_{26}\,,\right.\nonumber \\ 
& &\,\,\,\,\left.
M_{35}\mp iM_{36} \pm iM_{45}+M_{46}\,,
M_{13}\pm iM_{14} \mp iM_{23}+M_{24}\right\}=SU(3)\times U(1)\,.
\label{su3u1}
\ea
The explicit form of the $-1$ eigenspace is
\be
M^{(2)}=\left\{\left(\begin{matrix} A^{(2)} &0 \\ 0 & D^{(2)}\end{matrix}\right) | A^{(2)}=-KA^{(2)}K\,,D^{(2)}=-LD^{(2)}L\right\}\,,
\ee
where
\be
A^{(2)}=\sum_{\mu=0}^3a^{(2)}_\mu\gamma^{\mu 5}\,,\qquad D^{(2)}=\sum_{\mu=4}^9 d^{(2)}_\mu S^\mu\,,
\ee
with the $\gamma^{\mu 4}$ and $S^\mu$ defined in Appendix A

In terms of the $\Zop_4$-graded currents the GS action can be written as
\be
{\cal L}_{\mbox{\scriptsize GS}}=\int d^2\sigma\,\,
\sqrt{-g}g^{\alpha\beta}\Str(j^{(2)}_\alpha j^{(2)}_\beta)
+\epsilon^{\alpha\beta}\Str(j^{(1)}_\alpha j^{(3)}_\beta)\,.\label{z4gs}
\ee
When restricted to bosonic fields (by setting $j^{(1)}$ and $j^{(3)}=0$) the above action reduces to the 
sigma
model on $Sp(4)/SO(1,3)\times SO(6)/SU(3)\times U(1)$, in other words to  $AdS_4\times CP^3$. 
By construction this action has 24 supersymmetries, and has local $\kappa$ symmetry. We take this as
strong evidence that this action describes Type IIA string theory on the background~(\ref{bkd}).

\subsection{$\kappa$-symmetry}

Actions of the form~(\ref{z4gs}) constructed over cosets with a $\Zop_4$ automorphism typically have
$\kappa$-symmetry. The fermions in this theory live in the ${\bf 4}\times {\bf 6}$ of the $Sp(4)\times
SO(6)$ and so naively there are 24 real fermionic fields. In flat space and in $AdS_n\times S^n$ $\kappa$
symmetry can be used to gauge away half of them. In other, less conventional, models it has been recently
shown that $\kappa$ symmetry is trivial on-shell and cannot be used to gauge away any fermionic degrees of
freedom~\cite{bsll}. In the present model $\kappa$ symmetry will allow us to remove 8 real fermionic degrees
of freedom - leaving us with 16 physical fermions - the required number for Type IIA string theory on
$AdS_4\times CP^3$. In order to see that $\kappa$ symmetry can indeed be used to gauge away $1/3$ of the
fermions we consider for simplicity the superparticle.~\footnote{We defer a more detailed analysis of
$\kappa$-symmetry and gauge fixing to a forthcoming paper~\cite{bforth}.} In this case the action is simply
\be
\int d\tau\,\,
e\Str(j^{(2)}_\tau j^{(2)}_\tau)\,,
\label{supart}
\ee
where $e$ is the worldline metric. Picking the coset representatives as
\be
g=g_{\mbox{\scriptsize ferm}}\,\,g_{\mbox{\scriptsize bos}}\,,
\label{grpelt}
\ee
where $g_{\mbox{\scriptsize ferm}}$ ($g_{\mbox{\scriptsize bos}}$) is the exponential of
a purely fermionic (bosonic) Lie-algebra element, it is a simple exercise 
(see Appendix~\ref{appb}) to expand 
the above action to quadratic order in fermions. One finds that the two-fermion term is of the form
\be
{\cal L}_{\mbox{\scriptsize 2 ferm}}=\int d\tau e K^{ai\,,\,bj}\theta_{ai}\p_\tau\theta_{bj}=
\int d\tau e \sum_{\mu=0,\dots,3} p_\mu\gamma^{\mu 4}_{ac} \omega^{cb} \theta_{ai}\p_\tau\theta_{bi}
+\sum_{\mu=4,\dots 9} p_\mu S^\mu_{ij} \omega^{ab} \theta_{ai}\p_\tau\theta_{bj}\,,\label{2flag}
\ee
where $\mu=0,\dots,3$ and $\mu=4,\dots,9$ is the spacetime $AdS_4$ and $CP^3$ index with $p_\mu$ the spacetime momentum; $a,b,c=1,\dots,4$ and $i,j=1,\dots,6$ are $Sp(2)$ and $SO(6)$ indices; $\theta_{ai}$ are the 24 real fermionic fields in the coset~(\ref{coset}).
The on-shell condition comes from the variation of $e$ in the above action and is simply
\be
\Str(j^{(2)}_\tau j^{(2)}_\tau)=\eta^{\mu\nu}p_\mu p_\nu=0\,.
\label{onshell}
\ee 
We shown in appendix~\ref{appb} that on-shell $K_{ai\,,\,bj}$ has rank 16. This indicates that $\kappa$-symmetry of the 
action can be used to gauge away 8 real fermions leaving us with 16 physical fermions.

\subsection{Integrability}

The equations of motion that follow from~(\ref{z4gs}) are~\footnote{Note that since we are discussing 
classical integrability, the 
result of~\cite{BPR} is completely general for all actions of the 
form~(\ref{z4gs}) and does not depend on the specifics of the coset 
under investigation. This is reflected in the equations presented in this sub-section.}
\ba
0&=&\p_\a(\sqrt{-g}g^{\a\b}j^{(2)}_\b)-\sqrt{-g}g^{\a\b}\left[j^{(0)}_\a,j^{(2)}_\b\right]
+\frac{1}{2}\epsilon^{\a\b}\left(\left[j^{(1)}_\a,j^{(1)}_\b\right]-\left[j^{(3)}_\a,j^{(3)}_\b\right]
\right)\,,\label{eom1}\\
0&=&\left(\sqrt{-g}g^{\a\b}+\epsilon^{\a\b}\right)\left[j^{(3)}_\a,j^{(2)}_\b\right]\,,\label{eom2}\\
0&=&\left(\sqrt{-g}g^{\a\b}-\epsilon^{\a\b}\right)\left[j^{(1)}_\a,j^{(2)}_\b\right]\,.\label{eom3}
\ea
As is by now well known these equations follow from the zero curvature condition of a Lax connection~\cite{BPR}
(we follow the notation of~\cite{AAT})
\be
\p_\a{\cal L}_\b-\p_\b{\cal L}_\a-[{\cal L}_\a\,,\,{\cal L}_\b]=0\,,
\ee
where
\be
{\cal L}_\a=l_0 j^{(0)}_\a+l_1 j^{(1)}_\a+l_2\gamma_{\a\b}\epsilon^{\b\rho}j^{(2)}_\rho+l_3(j^{(1)}_\a+j^{(3)}_\a)
+l_4(j^{(1)}_\a-j^{(3)}_\a)\,.
\ee
The parameters $l_i$ are constants which depend on the spectral parameter $\lambda$ and are
\be
l_0=1\,,\qquad 
l_1=\frac{1+\lambda^2}{1-\lambda^2}\,,
l_2=\frac{2\lambda}{1-\lambda^2}\,,
\l_3=\pm\frac{1}{\sqrt{1-\lambda^2}}\,,l_4=\pm\frac{\lambda}{\sqrt{1-\lambda^2}}\,.
\ee
An infinite tower of local commuting charges may then be constructed using the monodromy matrix of the above Lax 
connection in the standard fashion.

\section{Conclusions}
\setcounter{equation}{0}

In this paper we have constructed the Green-Schwarz action for Type IIA string theory on $AdS_4\times CP^3$. This 
construction relies on the presence of a $\Zop_4$ automorphism for the coset~(\ref{coset}). We have argued that the 
action has an unconventional form of $\kappa$-symmetry which can be used to gauge-fix $1/3$ of the fermionic fields. 
We have also shown that the model is classically integrable. Since the construction is purely algebraic it may also 
be used to construct the GS action for the background $AdS_2\times CP^1$, in this case the coset being 
$OSp(2|3)/(SO(2)\times U(1))$. We expect that it should be possible to construct the corresponding Berkovits string 
for this background as well; which presumably will also be integrable. It would be interesting to see the precise 
form of the ghost sector of that theory. Given the classical integrability of this string theory and the 
proposed~\cite{m} dual descriptions in terms of three-dimensional  Chern-Simons theories coupled to matter it seems 
likely that these dualities may be investigated within the framework developed over the last few years in the study 
of $AdS_5\times S^5$ - ${\cal N}=4$ Super-Yang Mills duality~\cite{b}; indeed progress in this direction has already 
begun to appear~\cite{gtside}.

\section*{Acknowledgements}

I am grateful to Arkady Tseytlin and Chris Hull for a critical reading of the manuscript. I would also like to thanks John McGreevy for
many Bagger-Lambert sessions and Alessandro Torrielli for numerous discussions on integrability. This research is funded by EPSRC and
MCOIF.

\appendix

\section{Matrix representations of $Sp(2)$ and $SO(6)$}\label{appa}
\setcounter{equation}{0}

\subsection{The $Sp(2,R)\sim SO(2,3)$ algebra}

Using $4\times 4$-dimensional $\gamma$ matrices we may construct the $SO(2,3)$ Lie lie algebra. Indeed define
\be
\gamma^0 =  \left(\begin{matrix} 0 & i 1_2 \\ i 1_2 & 0 
\end{matrix}\right)\,,\qquad
\gamma^1 =  \left(\begin{matrix} 0 & i \sigma^1 \\ -i\sigma^1 & 0
\end{matrix}\right)\,,
\gamma^2 =  \left(\begin{matrix} 0 & i \sigma^2 \\ -i\sigma^2 & 0 
\end{matrix}\right)\,,\qquad
\gamma^3 =  \left(\begin{matrix} 0 & i \sigma^3 \\ -i\sigma^3 & 0
\end{matrix}\right)\,,
\ee
and $\gamma^4 = \gamma^0\gamma^1\gamma^2\gamma^3$. These satisfy the anti-commutation relations
\be
\{\gamma^\alpha\,,\,\gamma^\beta\}=2\eta^{\alpha\beta}\,,\qquad \alpha\,,\,\beta=0\,,\dots\,,4
\ee
where $\eta^{\alpha\beta}=\mbox{diag}(-1,1,1,1,-1)$. We may pick the charge conjugation matrix to be 
$C=\gamma^0\gamma^4$; the $\gamma$ matrices then satisfy
\be
\gamma^\alpha{}^\dagger=C^{-1}\gamma^\alpha C\,.
\ee
The Lie algebra $SO(2,3)$ is generated by 
\be
\gamma^{\alpha\beta}=\frac{1}{2}\left(\gamma^\alpha\gamma^\beta-\gamma^\beta\gamma^\alpha\right)\,,
\ee
with the $\gamma^{\alpha\beta}$ satisfying the reality condition
\be
\gamma^{\alpha\beta}{}^\dagger=-C^{-1}\gamma^\alpha C\,.
\ee
It is easy to check that the $\gamma^{\alpha\beta}$ satisfy the symplectic condition
\be
\gamma^{\alpha\beta}{}^t\omega+\omega \gamma^{\alpha\beta}=0\,,
\ee
showing that $SO(2,3)\sim Sp(2,R)$. We may also easily show that $G$ the $\Zop_4$ automorphism defined 
in section 2 acts as $+1$ on the $SO(1,3)$ subgroup generated by$\gamma^{\alpha\beta}$ with 
$\alpha,\beta=1,\dots,4$.

\subsection{The $S0(6)\sim SU(4)$ Lie algebra}

The Lie algebra $SO(6)$ contains an $SU(3)\times U(1)$ subalgebra (see equation~(\ref{su3u1})). We label 
the remaining 6 generators as
$S^\mu$, $\mu=4\,,\dots\,,9$. A convenient parametrisation of these is given by
\ba
S^4&=&M_{14}+M_{23}\,,\qquad S^5=M_{24}-M_{13} \,,\qquad S^6=M_{16}-M_{25}\,,\nonumber \\
S^7&=&M_{26}-M_{15}\,,\qquad S^8=M_{36}-M_{45} \,,\qquad S^9=M_{46}-M_{35}\,.
\ea

\section{Fermionic quadratic terms}\label{appb}
\setcounter{equation}{0}

If we pick the group element $g$ as in equation~(\ref{grpelt}), then the superparticle action~(\ref{supart}) to quadratic order in fermionic fields is
\be
{\cal L}_{\mbox{\scriptsize 2 ferm}}=\int d\tau e\, \Str(g_{\mbox{\scriptsize 
bos}}j^{(2)}_{\mbox{\scriptsize bos}}g^{-1}_{\mbox{\scriptsize bos}}\Theta^t\p_\tau\Theta)\,,
\ee
where the 24 real fermions $\theta_{ai}$ are grouped into an $OSp(2|6)$ matrix
\be
\Theta=\left(\begin{matrix} 0 &\theta_{ai} \\ (\theta^t \omega)_{jb} &0 \end{matrix}\right)\,,
\ee
and $g_{\mbox{\scriptsize ferm}}=\exp(\Theta)$. Explicitly ${\cal L}_{\mbox{\scriptsize 2 ferm}}$ is given 
in equation~(\ref{2flag}) where
\be
p_\mu= \Str(g_{\mbox{\scriptsize bos}}j^{(2)}_{\mbox{\scriptsize bos}}g^{-1}_{\mbox{\scriptsize bos}}
\Gamma^\mu)/\Str(\Gamma^\mu \Gamma^\mu)
\ee
where
\be
\Gamma^\mu=\gamma^{\mu 4}\otimes 1\,,\,\,\, \mu=0\,,\dots\,,3\qquad\Gamma^\mu=\omega\otimes S^\mu\,,\,\,\,
\mu=4\,,\dots\,,9\,.
\ee
For generic on-shell momenta $p_\mu$ (\ie{} those that satisfy equation~(\ref{onshell})) the matrix 
$K_{ai\,,\,bj}$ 
may be brought into the form
\be
p_0\gamma^{04}_{ac} \omega^{cb} \otimes\delta_{ij}+p_5 \omega^{ac}\otimes S^5_{ij}\sim
\left(\begin{matrix} p_1 & 0 & 0 \\ 0 & p_5 & -p_1 \\ 0 & p_1 & -p_5 \end{matrix}\right)^{\oplus 8}
\,,
\ee
by a global $Sp(2)\times SO(6)$ rotation. Since the on-shell condition reduces to $p_5=\pm p_0$, we see that for generic momenta
$K_{ai\,,\,bj}$ has rank 16.

\end{document}